\documentclass[twocolumn, amsmath,trackchanges]{aastex702}
\reportnum{NORDITA-2025-039}

\usepackage{siunitx}
\usepackage{physics}

\begin{document}

\title{Nonlocal Evolution of Magnetic Helicity across the Magnetohydrodynamic--Kinetic Interface}

\author[orcid=0000-0002-3735-0616, gname='Masatomi', sname='Iizawa']{Masatomi Iizawa}
\affiliation{Institut f\"ur Theoretische Physik, Technische Universit\"at Braunschweig, Mendelssohnstr. 3, D-38106 Braunschweig, Germany}
\email[show]{masatomi.iizawa@tu-braunschweig.de}

\author[orcid=0000-0002-5332-8881, gname='Yasuhito', sname='Narita']{Yasuhito Narita}
\affiliation{Institut f\"ur Theoretische Physik, Technische Universit\"at Braunschweig, Mendelssohnstr. 3, D-38106 Braunschweig, Germany}
\affiliation{Max Planck Institute for Solar System Research, Justus-von-Liebig-Weg 3, D-37077 G\"ottingen, Germany}
\email{}

\author[orcid=0000-0001-6096-0220, gname='Tommaso', sname='Alberti']{Tommaso Alberti}
\affiliation{Istituto Nazionale di Geofisica e Vulcanologia, via di Vigna Murata 605, 00143 Rome, Italy}
\email{}

\author[orcid=0000-0002-1989-3596, gname='Stuart', sname='Bale']{Stuart D. Bale}
\affiliation{Space Sciences Laboratory, University of California, Berkeley, CA 94720-7450, USA}
\affiliation{Physics Department, University of California, Berkeley, CA 94720-7300, USA}
\email{}

\author[orcid=0000-0002-7304-021X, gname='Axel', sname='Brandenburg']{Axel Brandenburg}
\affiliation{Nordita, KTH Royal Institute of Technology and Stockholm University, 10691 Stockholm, Sweden}
\affiliation{The Oskar Klein Centre, Department of Astronomy, Stockholm University, 10691 Stockholm, Sweden}
\affiliation{School of Natural Sciences and Medicine, Ilia State University, 0194 Tbilisi, Georgia}
\affiliation{McWilliams Center for Cosmology and Department of Physics, Carnegie Mellon University, Pittsburgh, Pennsylvania 15213, USA}
\email{}

\author[orcid=0000-0002-8932-0793, gname='Abraham', sname='Chian']{Abraham C.-L. Chian}
\affiliation{School of Computer and Mathematical Sciences, Adelaide University, Adelaide, SA 5005, Australia}
\affiliation{National Institute for Space Research (INPE), P.O. Box 515, S\~{a}o Jos\'{e} dos Campos, S\~{a}o Paulo 12227-010, Brazil}
\email{}

\author[orcid=0000-0002-5028-8482, gname='Horia', sname='Comisel']{Horia Comi\c{s}el}
\affiliation{Institute for Space Sciences, Atomi\c{s}tilor 409, P.O. Box MG-23, Bucharest-M\v{a}gurele, RO-077125, Romania}
\email{}

\author[orcid=0000-0002-4784-0301, gname='Shuichi', sname='Matsukiyo']{Shuichi Matsukiyo}
\affiliation{Faculty of Engineering Sciences, Kyushu University, Kasuga, Fukuoka 816-8580, Japan}
\affiliation{Quantum and Spacetime Research Institute (QuaSR), Kyushu University, Motooka, Nishi-ku, Fukuoka 819-0395, Japan}
\affiliation{International Research Center for Space and Planetary Environmental Science (i-SPES), Kyushu University, Motooka, Nishi-ku, Fukuoka 819-0395, Japan}
\affiliation{Institute of Laser Engineering, Osaka University, Suita, Osaka 565-0871, Japan}
\email{}

\author[orcid=0000-0002-5242-7634, gname='Nobumitsu', sname='Yokoi']{Nobumitsu Yokoi}
\affiliation{Department of Earth and Planetary Science, University of Tokyo, 7-3-1 Hongo, Bunkyo, Tokyo 113-0033, Japan}
\affiliation{Institute of Fluid Science, Tohoku University, 2-1-1, Katahira, Aoba-ku, Sendai 980-8577, Japan}
\affiliation{Laboratoire de Mécanique des Fluides et d'Acoustique (LMFA), UMR5509, Université de Lyon, École Centrale de Lyon, CNRS, Université Claude Bernard Lyon 1, INSA Lyon, 36 Avenue Guy de Collonge, 69130 Écully, France}
\affiliation{Nordita, KTH Royal Institute of Technology and Stockholm University, 10691 Stockholm, Sweden}
\email{}

\begin{abstract}
Magnetic helicity is typically indistinguishable from noise in the solar wind inertial range. Using Parker Solar Probe measurements (0.1--0.8 au), we report a robust bihelical structure emerging at the magnetohydrodynamic (MHD)--kinetic interface. This structure, formed by a striking pair of positive and negative peaks, exhibits a systematic spectral migration toward lower frequencies with increasing heliocentric distance. Crucially, this evolution deviates from local kinetic scaling laws, such as the radial decline of the ion cyclotron frequency, reflecting a nonlocal evolution of magnetic topology. Our findings provide a new observational constraint on both nonlocal MHD transport processes and the nature of the helicity barrier in the kinetic regime.
\end{abstract}



\section{Introduction}
The solar wind provides a dynamic laboratory for studying the transition from magnetohydrodynamic (MHD) to kinetic plasma turbulence \citep{Bruno2013}. A fundamental quantity in this transition is magnetic helicity, which characterizes the topological complexity of magnetic fields. While MHD theory treats helicity as a rugged invariant \citep{Matthaeus1982_jgr,Matthaeus1982_prl,Brown1999}, its fate at the ion kinetic scales, where fluid descriptions fail and wave-particle interactions dominate, remains a frontier of plasma physics \citep{Bowen2020}. Traditionally, helical fluctuations at sub-ion scales have been interpreted as local wave modes, such as ion cyclotron waves (ICWs) or kinetic Alfvén waves (KAWs), governed by the local plasma environment \citep{Leamon1998, Podesta2013}. However, recent theoretical developments, most notably the helicity barrier \citep{Meyrand2021, Squire2022}, suggest that the interplay between magnetic helicity and kinetic damping can fundamentally alter the energy cascade, potentially leading to the accumulation or redistribution of helical signatures across scales.

Magnetic helicity is often dismissed as stochastic noise within the MHD inertial range \citep{Matthaeus1982_prl, Smith2003}. Consequently, the recent identification of helical peaks near ion kinetic scales was a major surprise \citep{Leamon1998, Podesta2013, Telloni2015, Alberti2022}. These signatures, however, have remained snapshots confined to specific local plasma conditions. Tracking their radial evolution was impossible for earlier missions like Helios \citep{Matthaeus1982_jgr} and Voyager \citep{Bruno1986}, which lacked the sampling rates to probe the sub-ion regime above $10^{-2}$ Hz. It remains a fundamental missing link whether these structures are merely transient local fluctuations or part of a broader, non-adiabatic transformation. In this letter, we exploit the high-resolution magnetic field measurements obtained by the Parker Solar Probe (PSP) during its deep penetration into the inner heliosphere. By analyzing the magnetic helicity spectra at frequencies above 1 Hz, we identified robust bi-helical structures that migrate systematically  across heliocentric distances. Our study primarily focuses on the quantified discrepancy between the observed drift and the local kinetic scaling laws, suggesting a fundamentally non-local evolution of magnetic topology at the MHD-kinetic interface.

\section{Methodology and its validity}
We estimated the magnetic helicity density spectrum \citep{Matthaeus1982_prl, Matthaeus1982_jgr, Narita2009} using data from the FIELDS fluxgate magnetometer \citep{Bale2016} on Parker Solar Probe (PSP) \citep{Fox2016}. We focused on data collected from September 29, 2021, 00:00:00 UTC, to November 22, 2021, 00:00:00 UTC, a period characterized by short sampling intervals and a wide radial swing from 0.100 AU to 0.782 AU in 0.002 AU increments. For each increment, one hour of data was collected, with sampling intervals of $\SI{3441}{\micro s}$ (0.100--$\SI{0.160}{AU}$), $\SI{6827}{\micro s}$ (0.162--$\SI{0.260}{AU}$), and $\SI{109227}{\micro s}$ (0.262--$\SI{0.782}{AU}$), which affect the acquirable frequency bands. Cubic spline interpolation addressed $\sim\SI{1}{\micro s}$ fluctuations in sampling intervals. The data with inconsistent sample counts per hour were treated as missing, especially beyond $\SI{0.434}{AU}$. The reduced normalized magnetic helicity density spectra $\sigma_{\mathrm{m}}$ \citep{Matthaeus1982_prl, Matthaeus1982_jgr, Narita2023}, defined as
\begin{align}
\sigma_{\mathrm{m}} = \frac{2\Im \left[ \ev{b^{\ast}_{\mathrm{T}}b_{\mathrm{N}}} \right]}{\ev{|b_{\mathrm{T}}|^{2}} + \ev{|b_{\mathrm{N}}|^{2}}}
\end{align}
were calculated using magnetic fields $b_{\mathrm{T}}$ and $b_{\mathrm{N}}$ in the RTN coordinate system in the frequency domain, which are transformed using fast Fourier transform (FFT), i.e. $b_{\mathrm{T}}(f) = \sum^{L-1}_{t=0}\exp (-2\pi j f t / L)B_{\mathrm{T}}(t)$ where $L$ is the number of samples in the time domain, and $B_{\mathrm{T}}(t)$ and $b_{\mathrm{N}}(f)$ represent the magnetic field in the time and frequency domain, respectively. The normal component $b_{\mathrm{N}}(f)$ is defined analogously. $\ev{|b_{\mathrm{T}}|^{2}}$ and $\ev{|b_{\mathrm{N}}|^{2}}$ represent power spectral density. This definition yields a positive sign for helicity when it is right-handed with respect to the radial outward direction, consistent with the definition of the cross-spectral density, for example, \textit{complex coherence} in signal processing \citep{Ramrez2022}, though opposite to \citet{Podesta2011}. Spectral density estimation employed Welch's method with a Hanning window \citep{Welch1967}. It is important to emphasize that $\sigma_{\mathrm{m}}$ is a reduced quantity derived from single-spacecraft time series under Taylor’s frozen-in hypothesis (TH) \citep{Taylor1938}. This metric $\sigma_{\mathrm{m}}$ specifically measures the sense of magnetic field rotation (polarization) around the radial sampling direction \citep{Telloni2015}. Consequently, due to the inherent limitations of single-point measurements, this approach does not capture the full three-dimensional helicity, as it excludes fluctuations and wavevectors associated with the radial component \citep{Matthaeus1982_jgr, Narita2009}.

\begin{figure}[t]
\plotone{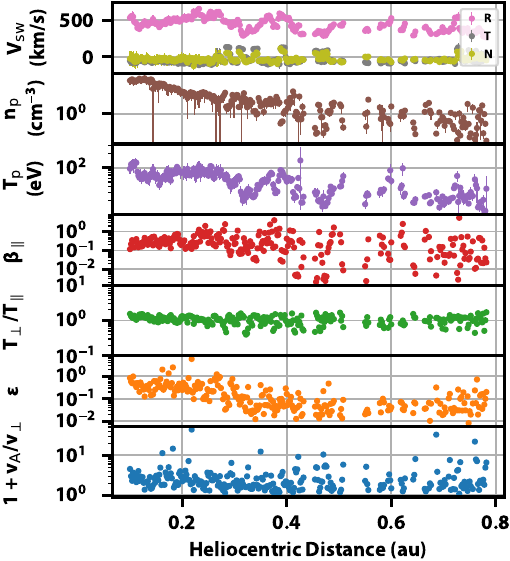}
\caption{Radial evolution of plasma parameters and Taylor hypothesis diagnostics. Panels display (from top to bottom): solar wind flow speed $V_{\mathrm{sw}}$, proton density $n_{\mathrm{p}}$, proton temperature $T_{\mathrm{p}}$, parallel proton beta $\beta_{\parallel}$, and temperature anisotropy $T_{\perp} / T_{\parallel}$. The bottom panels show the sweeping parameter $\epsilon$ and the Doppler shift factor $1+V_{\mathrm{A}}/V_{\perp}$ (see text for definitions), used to assess the validity of the Taylor frozen-in hypothesis. All parameters are averaged over one-hour intervals. Missing data points are omitted from the plots.
\label{fig:flow_speed_etc}}
\end{figure}

To assess the validity of TH used for the wavenumber transformation, we evaluate the effects of both random sweeping and wave propagation in the plasma frame. Fig.~\ref{fig:flow_speed_etc} displays the radial evolution of the dimensionless parameter $\epsilon = \delta u_{0}/(\sqrt{2} V_{\perp})$ and the Doppler shift factor assuming Alfvén waves $1+V_{\mathrm{A}}/V_{\perp}$, where $V_{\perp}$ is the spacecraft velocity component perpendicular to the local magnetic field in the plasma frame \citep{Perez2021}. The parameter $\epsilon$ quantifies the ratio of the velocity fluctuations at the energy-containing scale $\delta u_{0}$ to the sampling speed. As shown in Fig.~\ref{fig:flow_speed_etc}, $\epsilon$ decreases from $\approx 0.3$ near the perihelion ($\approx 0.1\,\mbox{AU}$) to $\approx 0.03$ at larger distances ($\approx 0.7\,\mbox{AU}$). This radial trend is in excellent agreement with theoretical predictions for the inner heliosphere \citep{Bourouaine2019}. We used the standard deviation of velocity fluctuations for $\delta u_{0}$ without applying the moving average filter used by \citet{Perez2021}. Since our method includes fluctuations from all measured scales, the $\epsilon$ values in Fig.~\ref{fig:flow_speed_etc} represent a conservative upper bound. The actual sweeping effect is likely smaller, reinforcing the validity of the TH as $\epsilon$ decreases to $\ll 1$ with distance. In the near-Sun region, the relatively high $\epsilon$ values ($\gtrsim 0.5$) and the low Alfvén Mach number regime suggest that the strict frozen-in condition may be violated, potentially subjecting the frequency-to-wavenumber mapping to spectral broadening. However, theoretical models and numerical simulations demonstrate that even under strong sweeping (large $\epsilon$), the spectral index of the frequency spectrum remains identical to that of the underlying wavenumber spectrum \citep{Wilczek2012, Bourouaine2018}. Thus, our estimates of magnetic helicity spectral slopes remain robust against the deviations from TH.
We note that the plasma parameters were obtained from PSP's SPAN-I \citep{Livi2022}. While partial field-of-view obstruction may cause underestimation of proton moments during radial flow \citep{Livi2022}, such systematic offsets act primarily as linear scaling factors and cannot reproduce the observed order-of-magnitude radial migration addressed in this study.
Regarding spatial scales, the factor $1+V_{\mathrm{A}}/V_{\perp}$ fluctuates between 1 and 4, with a mean of approximately 2.5. This indicates that due to the non-negligible phase speed of Alfvénic fluctuations in low-Mach-number flows \citep{Klein2014}, the derived wavenumbers may be overestimated (and physical scales underestimated) by a factor of $\sim 2.5$. Nevertheless, as we investigate spectral features spanning multiple orders of magnitude, this factor-of-few shift is negligible relative to the broad spectral evolution. Furthermore, this implies that changes in the sampling angle $\theta_{BV}$ cannot drive order-of-magnitude systematic trends, as its effect via $V_{\perp}$ is secondary.

The radial evolution of magnetic helicity signatures can originate from several non-local processes, such as non-linear mode coupling across distinct scales, large-scale advection, or the emergence of a helicity barrier. In the classical framework of MHD turbulent transport, these processes are typically quantified by the time evolution of the magnetic helicity density spectrum $H_{M}(k,t)=\Re \oint_{|\boldsymbol{k}|=k} \hat{\boldsymbol{A}}_{\boldsymbol{k}}\cdot \hat{\boldsymbol{B}}_{\boldsymbol{k}}^{\ast} d\Omega(\boldsymbol{k})$ where $d\Omega(\boldsymbol{k}) = k^2 \sin\theta d\theta d\phi$. That is,
\begin{align}
\frac{\partial}{\partial t} H_{M} (k, t) = 2 S_{M}(k, t)-2\eta k^{2} H_{M}(k, t)
\end{align}
where $S_{M}(k,t)$ is the transfer function defined as $\Re \oint_{|\boldsymbol{k}|=k} \hat{\mathcal{E}}_{\boldsymbol{k}}\cdot \hat{\boldsymbol{B}}_{\boldsymbol{k}}^{\ast} d\Omega(\boldsymbol{k})$, using the electromotive force $\hat{\mathcal{E}}_{\boldsymbol{k}} = \widehat{\boldsymbol{u}\times\boldsymbol{B}}$, and magnetic diffusivity $\eta$ \citep{Brandenburg2001, Mueller2012}. The spectral flux of magnetic helicity is defined as
\begin{align}
G_{M}(k,t) = \int^{\infty}_{k} 2 S_{M} (k', t) dk'.
\end{align}
Because $S_{M}$ is the third-order moment of the velocity field $\boldsymbol{u}$ and the magnetic field $\boldsymbol{B}$, its direct computation requires high-sampling rate data on the velocity field $\boldsymbol{u}$. However, in tenuous plasma environments, particle instruments must scan energy and direction, which fundamentally limits the sampling rate; consequently, obtaining velocity spectra at or above the order of 1~Hz remains unfeasible \citep{Wilson2022, Livi2022, Case2020}. Alternatives to this direct approach also face significant hurdles. Using the fluctuating electric field $\boldsymbol{E}$ as a proxy for the electromotive force $\boldsymbol{u} \times \boldsymbol{B}$ is unreliable below 100 Hz due to antenna characteristics and the plasma environment around the spacecraft \citep{Podesta2013, Imachi2007, Hartley2017, Karbashewski2023}. Furthermore, approximating $\boldsymbol{E}$ using the generalized Ohm’s law within the framework of Electron MHD (EMHD) requires the current density $\boldsymbol{j} = \nabla \times \boldsymbol{B}$, which necessitates multi-point spatial derivatives which are inaccessible to single-spacecraft measurements \citep{Narita2009, Glassmeier2001}. Given these constraints, a direct calculation of $S_{M}$ in the region of interest is currently impossible. Furthermore, the applicability of this MHD-based description to the MHD-kinetic interface remains theoretically unverified. Regardless of the underlying framework, the systematic spectral migration of helicity signatures reflects the integrated effect of all local and non-local transport processes.

\section{Results.}
\begin{figure}[t]
\plotone{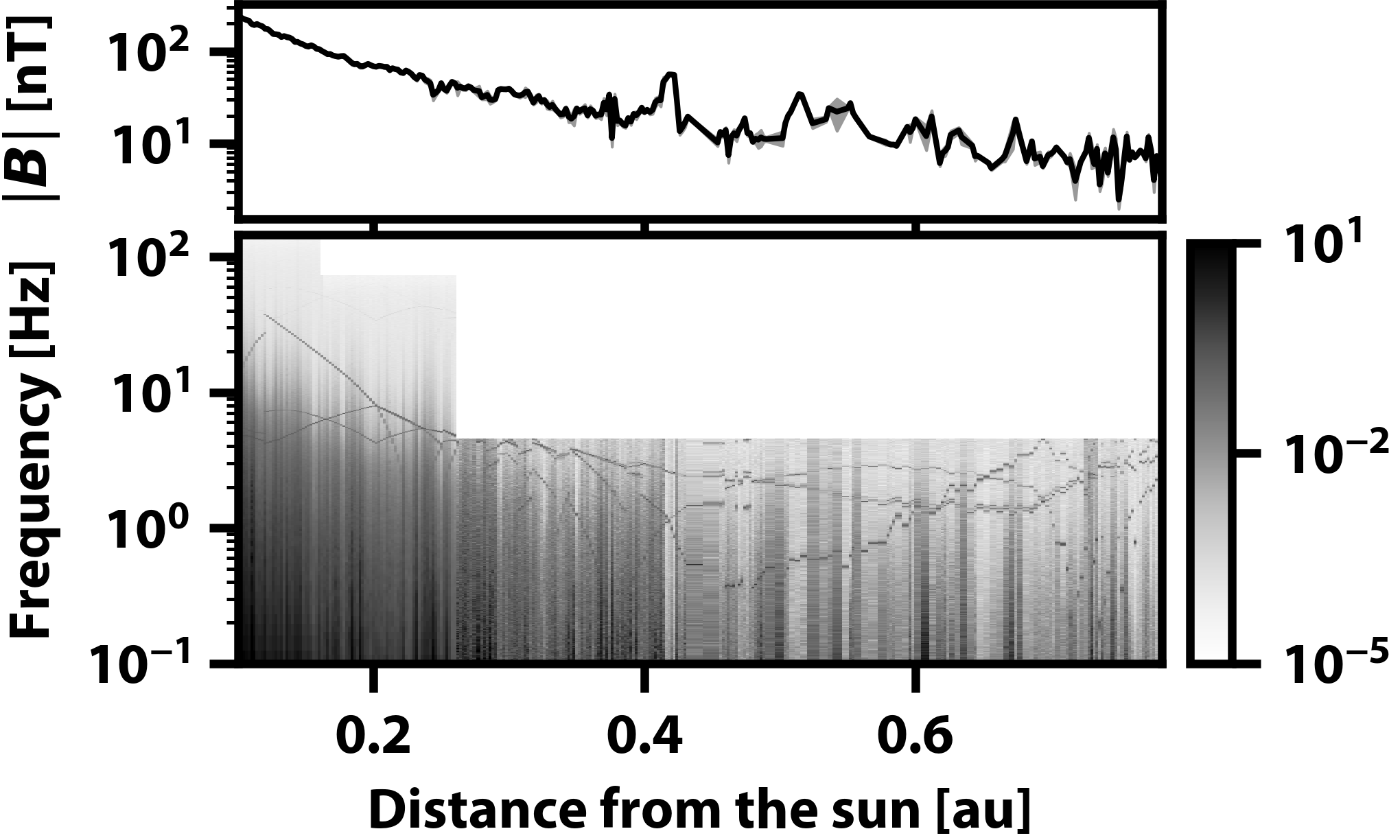}
\plotone{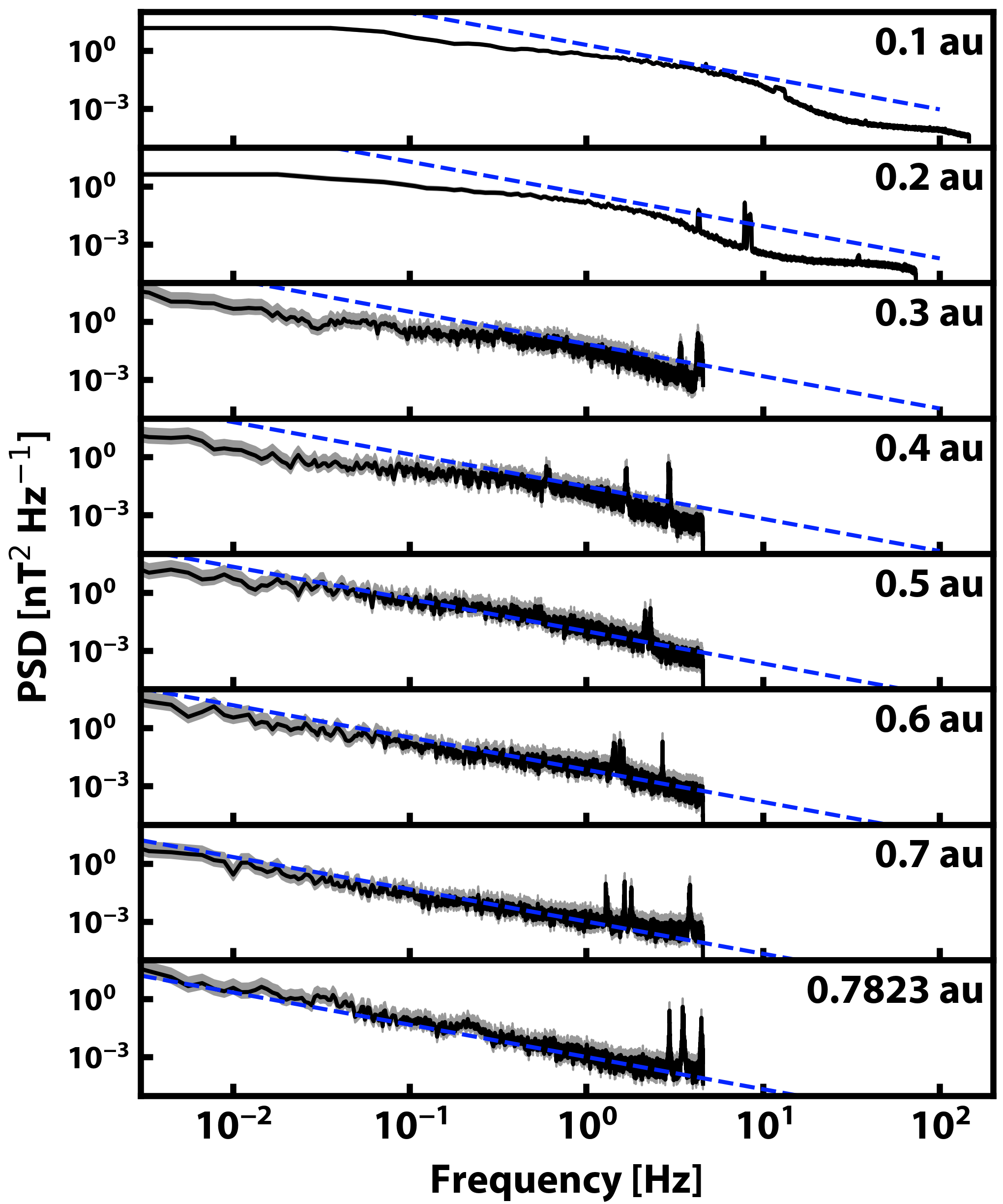}
\caption{Radial evolution of magnetic field intensity and power spectral density. [Top] Magnetic field magnitude $|\boldsymbol{B}|$ as a function of heliocentric distance. The gray shaded region represents the standard deviation calculated over the one-hour intervals. [Middle] PSD of magnetic field fluctuations shown as a color map in the frequency–distance plane. The spectra are computed via Welch’s method with a segment length of $2^{13}$ points. [Bottom] Representative 1D spectra at seven heliocentric distances, extracted from the middle panel. Blue dashed lines denote a $-5/3$ spectral index for reference. Deviations from this scaling are visible as the spacecraft approaches the Sun.
\label{fig:PSD_au}}
\end{figure}

The magnetic field power spectral density (PSD) shown in Fig.~\ref{fig:PSD_au} consistently exhibits an inertial subrange characterized by a Kolmogorov $-5/3$ energy decay, aligning with previous observations \citep{Podesta2013, Chen2020, Alberti2022}. Notably, at 0.1 AU, our high-sampling-rate measurements extended the observable frequency band by approximately two orders of magnitude compared to conventional datasets. However, the dissipation range remains uncaptured, suggesting that the Kolmogorov scale exceeds $\SI{2e2}{Hz}$. While previous studies have reported that the spectral index approaches $-3/2$ as heliocentric distance decreases, with a significant transition occurring around 0.4--0.6~AU \citep{Chen2020}, our statistical analysis generally supports this trend. It is worth noting, however, that the radial evolution of turbulence can be sensitive to the sampling method. While statistical surveys across diverse solar wind streams typically show this spectral evolution, \citet{Perrone2022} observed a more stable spectral slope when tracking a single plasma parcel from the same coronal source. Although our study adopts a statistical approach consistent with \citet{Chen2020}, resolving the discrepancy between these Eulerian and Lagrangian perspectives remains beyond the scope of this paper. Proposed mechanisms for this radial transition include the breakdown of self-similarity (a monofractal to multifractal transition) \citep{Alberti2020} and the evolving coupling between velocity and magnetic field fluctuations \citep{Alberti2022_ApJL}.

\begin{figure}[t]
\plotone{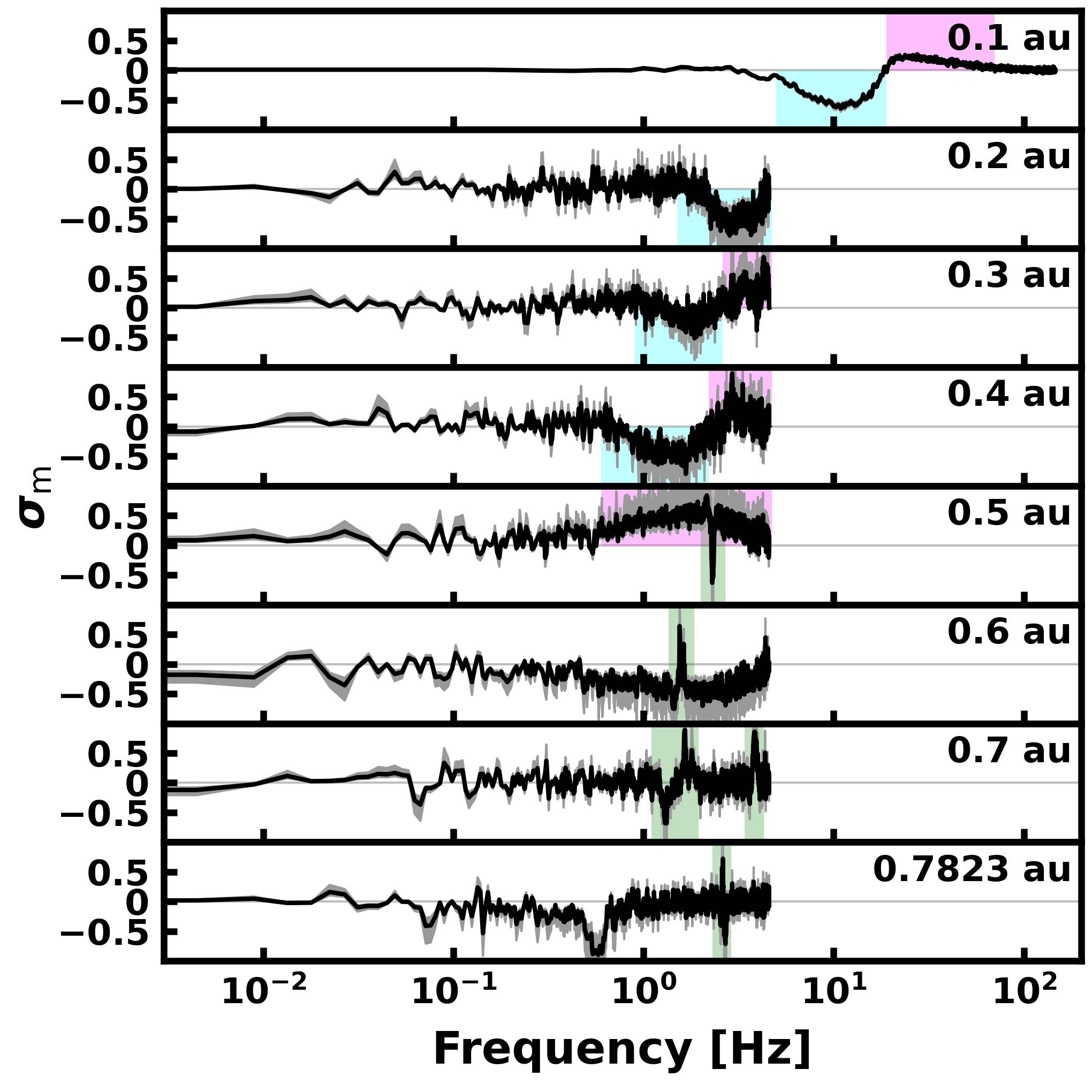}
\caption{Normalized magnetic helicity density spectra $\sigma_{\mathrm{m}}$ in the spacecraft frame. The spectra are computed via Welch’s method (segment length $2^{13}$), with gray shaded regions indicating 95\% confidence intervals. Red and blue areas represent positive and negative helicity, respectively, while green areas denote suspected instrumental artifacts. Note that individual intervals may capture periods of transiently low wave amplitude due to the inherent intermittency of solar wind turbulence \citep{Bruno2013, Jagarlamudi2021, Froment2023}. The comprehensive evolution of these modes across the inner heliosphere is further detailed in Fig.~\ref{fig:au_helicity_normalized}.
\label{fig:helicity_normalized}}
\end{figure}

\begin{figure}[t]
\plotone{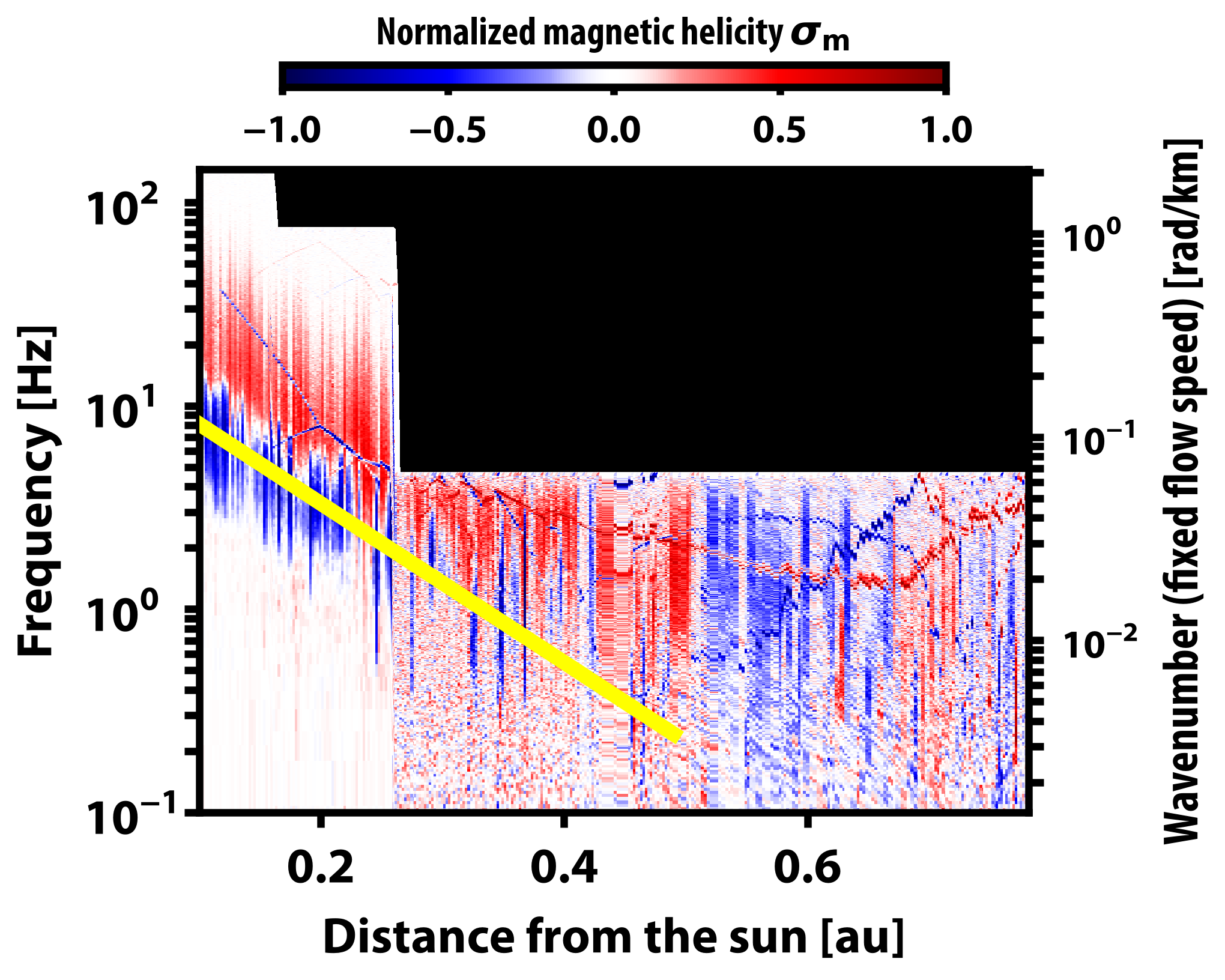}
\plotone{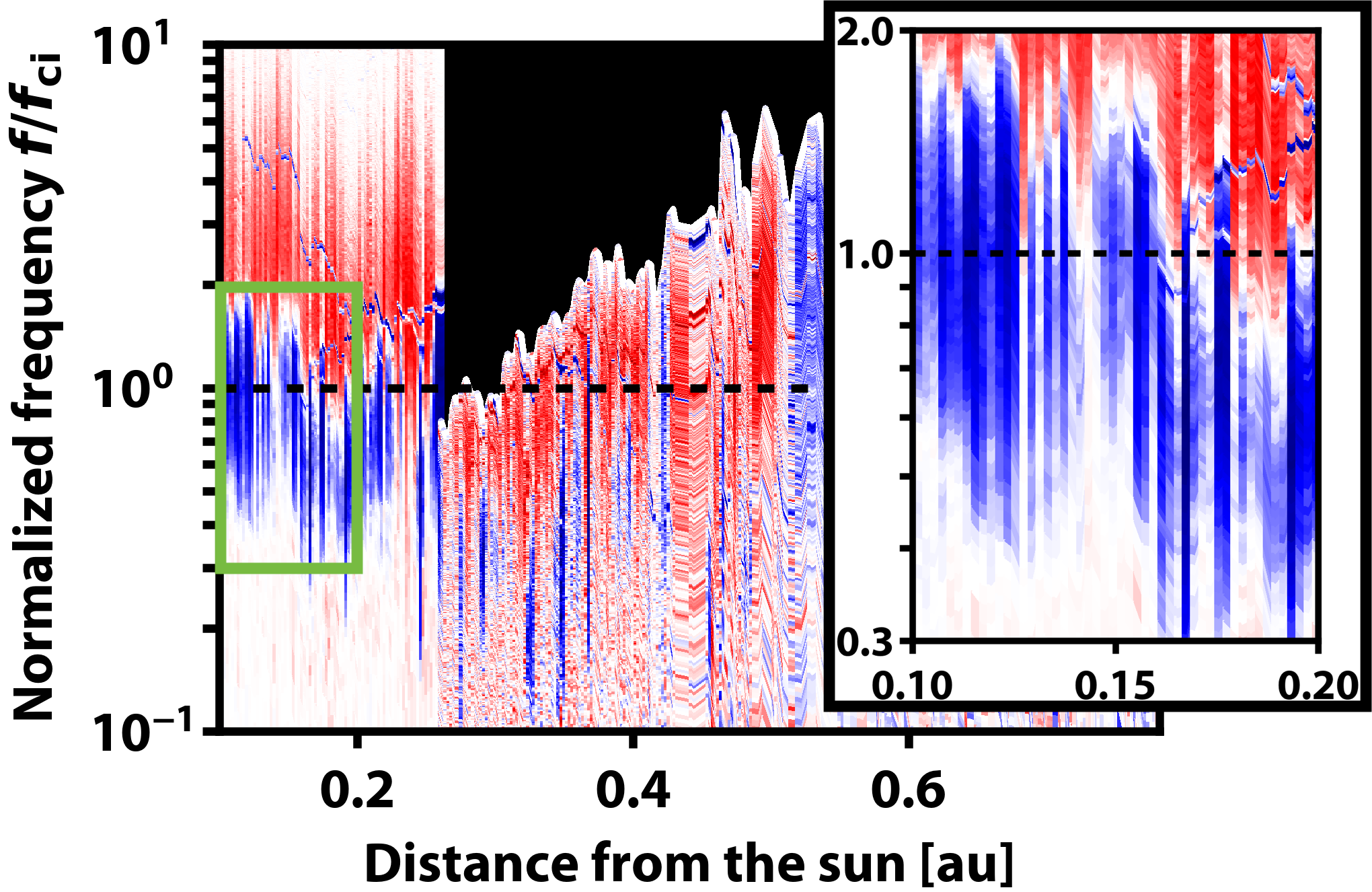}
\plotone{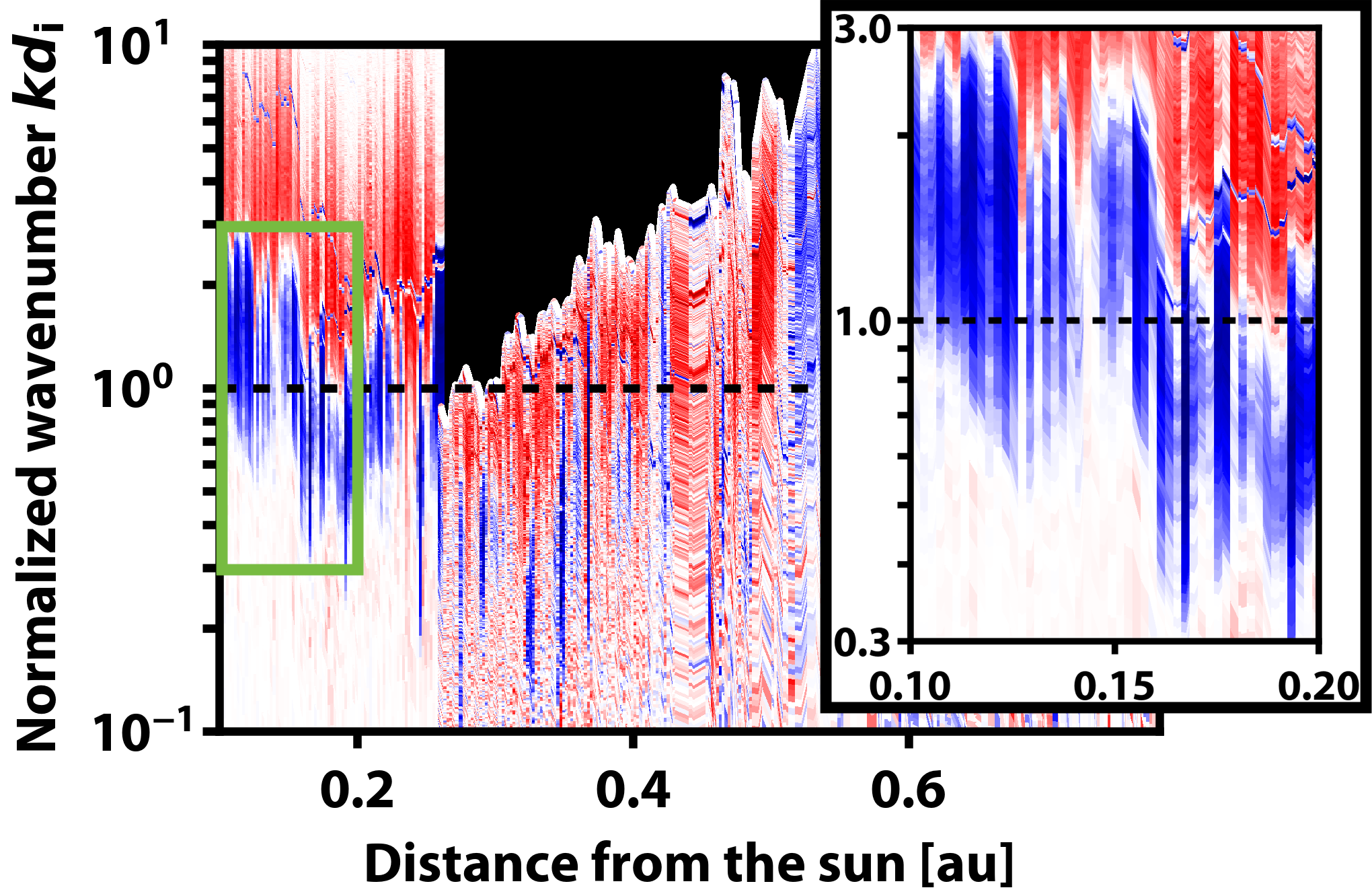}
\caption{Radial evolution of magnetic helicity and characteristic scale scaling. [Top] Color map of the reduced normalized magnetic helicity density $\sigma_{\mathrm{m}}$ in the frequency–distance plane. Black regions indicate data gaps due to low sampling rates. The yellow line is $f = f_0 \exp(-\lambda r)$, where $f_0 = 20$ Hz and $\lambda = 9\,\text{AU}^{-1}$ serve as a rough guide for the observed spectral migration. The right-hand axis provides an indicative wavenumber scale based on a constant solar wind speed $V_{\mathrm{sw}} = 441.4\,\text{km/s}$ (the average of the moment data); this axis serves as an order-of-magnitude reference. [Middle] Spectra normalized by the local proton cyclotron frequency, $f/f_{\mathrm{ci}}$. The inset shows an enlarged view of the green-outlined area. [Bottom] Spectra normalized by the ion inertial length, $k d_{\mathrm{i}}$. Despite the scatter induced by variable local conditions, the helicity peaks in the normalized plots do not collapse onto a fixed scale (e.g., $f/f_{\mathrm{ci}} \sim \mbox{const.}$ or $k d_{\mathrm{i}} \sim \mbox{const.}$). Instead, they exhibit a systematic migration toward larger scales relative to the local ion scales as distance increases. 
\label{fig:au_helicity_normalized}}
\end{figure}

More significantly, the normalized magnetic helicity density spectra, as exemplified in Fig.~\ref{fig:helicity_normalized}, reveal a prominent negative peak at \SI{10}{Hz} and a positive peak at \SI{20}{Hz} at \SI{0.1}{AU}. The statistical significance of these peaks is confirmed by 95\% confidence intervals that exclude zero. As the heliocentric distance increases, these peaks continuously shift to lower frequencies. This trend is clearly depicted in the top figure of Fig.~\ref{fig:au_helicity_normalized}. This shift, observed across 0.1--\SI{0.4}{AU}, provides a characteristic observational signature of the systematic growth of helical structures toward larger spatial scales. While some peaks appear to vanish at intermediate distances (e.g., 0.2~AU), this is an observational artifact caused by lower sampling rates in survey mode rather than a physical disappearance of the mode; these peaks re-emerge in the observable band beyond 0.3~AU, exhibiting a continuous shift toward lower frequencies. Beyond approximately \SI{0.5}{AU}, this clear helicity structure becomes less discernible. Minor fluctuations in spectral shift reflect the influence of transient events, as measurements across different distances are not concurrent. While some sharp, narrow peaks are considered instrumental artifacts, and others may capture solar events spanning several weeks, these factors do not alter our fundamental conclusions. This study does not aim to prove the ubiquity of these phenomena in the inner heliosphere, but rather to demonstrate their feasibility and empirical occurrence.

In Fig.~\ref{fig:au_helicity_normalized}, the middle and bottom panels display the spectra normalized by the local proton cyclotron frequency $f_{\mathrm{ci}}$ and the wavenumber scaled by the ion inertial length $k\,d_{\mathrm{i}}$, respectively. These normalizations serve to test whether the characteristic scales of magnetic helicity are governed primarily by local plasma instabilities. If the helical structures were generated and sustained purely by local kinetic processes, such as ion cyclotron or kinetic Alfvén instabilities observed in previous studies \citep{He2011, Huang2020,Bowen2020, Duan2021}, the spectral peaks would be expected to collapse onto a constant characteristic scale (e.g., $f/f_{\mathrm{ci}}\sim \mbox{const.}$ or $k\,d_{\mathrm{i}} \sim  \mbox{const.}$) regardless of the heliocentric distance. However, as clearly visualized in the enlarged insets of Fig.~\ref{fig:au_helicity_normalized}, the helical structures do not remain horizontally aligned with these local kinetic scales. Instead, the peaks systematically cross the constant-scale reference lines ($f/f_{\mathrm{ci}} = 1$ and $kd_{\mathrm{i}} = 1$) and exhibit a systematic migration toward larger scales. 

While a unified turbulent diffusion scale (e.g., $k(\rho_{\mathrm{p}} + d_{\mathrm{i}}) \sim 1$; \citet{Duan2020_ApJ, Duan2020_ApJS}) or the boundary of the turbulence ion diffusion range (T-IDR) \citep{He2020} provides a comprehensive physical boundary incorporating both dissipation and dispersion at the MHD-kinetic transition, we retain normalizations by individual parameters. The unified scale is derived from the parallel cyclotron resonance condition, and applying it to highly anisotropic solar wind turbulence introduces geometrical ambiguities. Furthermore, since the normalizations by both $f_{\mathrm{ci}}$ and $d_{\mathrm{i}}$ independently exhibit the same exponential spectral migration, the combined local scales would inevitably show an identical systematic slippage. This slippage of the energy-containing scale relative to the characteristic kinetic scales provides a compelling observational signature that the helical structures are not merely static local features but are dynamically evolving.

\section{Discussion and conclusion}
To properly interpret the observed dynamics of magnetic helicity, it is crucial to distinguish between the local injection (the seed) of helical fluctuations and their subsequent non-local evolution (transport). To understand the origin of the distinct positive and negative helicity peaks observed between 0.1 and 0.4 AU, the most compelling seed mechanism is the excitation by proton temperature anisotropy \citep{Podesta2011}. Specifically, pressure anisotropy ($T_{\mathrm{p}\perp}/T_{\mathrm{p}\parallel} > 1$) triggers left-hand polarized ion-cyclotron waves, while the firehose instability ($T_{\mathrm{p}\perp}/T_{\mathrm{p}\parallel} < 1$) excites right-hand polarized whistler waves. While the dispersion relation of L-mode waves exhibits $\omega$-axis symmetry in the plasma frame, the Doppler shift in the spacecraft frame reduces the angular frequency $\omega$ for wavevectors $k < 0$ relative to $k > 0$, naturally producing the observed split in magnetic helicity signs. This interpretation is further supported by the fact that R-mode whistler waves typically emerge at smaller scales than L-mode waves. Conversely, alternative mechanisms such as parametric decay \citep{DelZanna2001, Tenerani2013, Bruno2013, Bowen2018}, though capable of increasing magnetic compressibility \citep{Chen2020}, would likely result in random mixing rather than the clearly separated helicity components observed here; furthermore, simulations indicate that Alfvén waves subject to parametric breakdown diminish significantly within a few solar radii \citep{Tenerani2013}. Similarly, while mean-field dynamo theory predicts bi-helical magnetic helicity \citep{Blackman2003}, it assumes hemispheric sign reversals at scales far exceeding those in our study (2--\SI{25}{\micro Hz}) \citep{Brandenburg2011}, and the required latitudinal dependence remains unverifiable with current PSP data, awaiting future multi-spacecraft observations. Furthermore, recent observations by \citet{Xie2026} provide another compelling candidate for the local seeding of such bi-helical signatures. They demonstrated that at the MHD-kinetic interface, quasi-perpendicular kinetic Alfvén waves transfer energy to particles, which subsequently inject part of that energy into quasi-parallel ion cyclotron waves.

Once injected, the dynamic non-local evolution of magnetic helicity is further corroborated by the specific functional form of the observed frequency shift. While adiabatic solar wind expansion can increase turbulence scales \citep{Tu1993, Marsch1993, Grappin1996, Goldstein1999}, simple expansion models based on the frozen-in hypothesis ($f \sim U_{\mathrm{sw}}/d_{\mathrm{i}}$) and Parker's density profile ($n \propto r^{-2}$) predict a power-law shift of $f \propto 1/r$. This fundamentally differs from our observed exponential decay, $f \propto \exp(-\lambda r)$, ruling out passive geometric expansion as the primary driver and pointing instead to nonlinear fluid interactions. Specifically, the accumulation of magnetic helicity at scales exceeding local ion scales aligns with the \textit{helicity barrier} mechanism proposed for imbalanced kinetic Alfvén wave turbulence \citep{Meyrand2021, Squire2022}. This theory suggests that the transition from MHD to kinetic scales inhibits the forward energy cascade for high cross-helicity flows, forcing an inverse transfer and a pile-up of magnetic helicity at scales larger than the ion gyroradius. Consequently, the observed slippage of the helicity peak relative to ion scales (Fig.~\ref{fig:au_helicity_normalized}, bottom) provides a compelling observational signature of this barrier preventing efficient dissipation of helical fluctuations. Although third-order scaling laws, such as Yaglom's law \citep{Wang2023}, offer rigorous tools for determining cascade directions \citep{SorrisoValvo2007, SorrisoValvo2023, Wu2022, Wu2023, Consolini2020}, their application is precluded by the aforementioned sampling limitations. Relying on second-order spectral evolution, we find that the observed exponential migration poses a new challenge for theoretical modeling, as it deviates from the power-law evolution predicted by idealized EMHD simulations \citep{Kim2015}. This discrepancy likely stems from the complex interplay between proton-scale dynamics and global expansion in the solar wind. Future studies should investigate how the contributions of individual terms in the MHD equations vary under these realistic conditions, potentially exploring the role of residual helicity, i.e., the difference between kinetic and current helicity, as suggested by residual energy analysis \citep{Yokoi2007}.

Finally, we address the observed disappearance of the magnetic helicity structure beyond \SI{0.5}{AU}. The proximity of PSP's orbit to the ecliptic plane (within $\SI{\pm 3.4}{\degree}$), combined with significant fluctuations in the magnetic field strength near 0.4--\SI{0.5}{AU}, suggests that the spacecraft frequently crosses the heliospheric current sheet (HCS) \citep{Wilcox1965}. Such interactions likely disrupt the transport of helical fluctuations, potentially causing the coherent helicity signatures to vanish. To determine whether this disappearance represents a genuine spatial evolution or merely an orbital artifact caused by the HCS, future studies should extend this analysis using complementary data at larger heliocentric distances and higher latitudes obtained by the Solar Orbiter. In addition, this disappearance coincides with a reported transition in the radial dependence of energy spectra between 0.4 and 0.6 AU \citep{Alberti2020, Alberti2022_ApJL}, marking the shift from predominantly magnetohydrodynamic turbulence near the Sun to a more hydrodynamic-like regime at larger distances. Critically, while traditional theories often neglect the nonlinear advection term when evaluating the influence of background magnetic fields \citep{Alberti2022_ApJL}, our results suggest that this term is indispensable. Moreover, the applicability of this MHD framework to the MHD-kinetic interface remains an open question. Our results demonstrate that the integrated effect of all underlying fundamental processes follows neither passive expansion nor local kinetic dissipation. Instead, the observed exponential migration serves as a quantitative constraint for future models to either incorporate kinetic-scale effects or extend current turbulence paradigms.

\begin{acknowledgments}
MI appreciates the insightful discussion with Prof. Uwe Motschmann (TU Braunschweig, Germany). We thank the anonymous referee for helpful comments and suggestions. The FIELDS experiment on the Parker Solar Probe spacecraft was designed and developed under NASA contract NNN06AA01C. We acknowledge the NASA Parker Solar Probe Mission and the SWEAP team led by J. Kasper for use of data. This work was funded by the Deutsche Forschungsgemeinschaft (DFG, German Research Foundation) -- 535057280 and also by the Japan Society for the Promotion of Science (JSPS) Grants-in-Aid for Scientific Research JP23K25895 and JP26K00767. We acknowledge support by the Open Access Publication Funds of the Technische Universit\"{a}t Braunschweig.

\vspace{2mm}\noindent
\emph{Software and Data Availability.}
The codes used for the analysis and the plotted data are available on \dataset[https://doi.org/10.5281/zenodo.22779133]{https://doi.org/10.5281/zenodo.22779133}
\end{acknowledgments}

%
\facility{PSP}

\software{SunPy \citep{sunpy2020,sunpy_zenodo}, Astropy \citep{astropy2013,astropy2018,astropy2022}, SpiceyPy \citep{spiceypy}, Polars \citep{polars}, SciPy \citep{scipy}, NumPy \citep{numpy}, Matplotlib \citep{matplotlib}
          }





\begin{thebibliography}{}
\expandafter\ifx\csname natexlab\endcsname\relax\def\natexlab#1{#1}\fi
\providecommand{\url}[1]{\href{#1}{#1}}
\providecommand{\dodoi}[1]{doi:~\href{http://doi.org/#1}{\nolinkurl{#1}}}
\providecommand{\doeprint}[1]{\href{http://ascl.net/#1}{\nolinkurl{http://ascl.net/#1}}}
\providecommand{\doarXiv}[1]{\href{https://arxiv.org/abs/#1}{\nolinkurl{https://arxiv.org/abs/#1}}}

\bibitem[{T. Alberti {et~al.}(2022{\natexlab{a}})Alberti, Benella, Consolini, Stumpo, \& Benzi}]{Alberti2022_ApJL}
Alberti, T., Benella, S., Consolini, G., Stumpo, M., \& Benzi, R. 2022{\natexlab{a}}, \bibinfo{title}{Reconciling Parker Solar Probe Observations and Magnetohydrodynamic Theory,} Astrophys. J. Lett., 940, L13, \dodoi{10.3847/2041-8213/aca075}

\bibitem[{T. Alberti {et~al.}(2020)Alberti, Laurenza, Consolini, Milillo, Marcucci, Carbone, \& Bale}]{Alberti2020}
Alberti, T., Laurenza, M., Consolini, G., {et~al.} 2020, \bibinfo{title}{On the Scaling Properties of Magnetic-field Fluctuations through the Inner Heliosphere,} Astrophys. J., 902, 84, \dodoi{10.3847/1538-4357/abb3d2}

\bibitem[{T. Alberti {et~al.}(2022{\natexlab{b}})Alberti, Narita, Hadid, Heyner, Milillo, Plainaki, Auster, \& Richter}]{Alberti2022}
Alberti, T., Narita, Y., Hadid, L.~Z., {et~al.} 2022{\natexlab{b}}, \bibinfo{title}{Tracking of magnetic helicity evolution in the inner heliosphere: A radial alignment study,} Astron. Astrophys., 664, L8, \dodoi{10.1051/0004-6361/202244314}

\bibitem[{A. {Annex} {et~al.}(2020){Annex}, {Pearson}, {Seignovert}, {Carcich}, {Eichhorn}, {Mapel}, {von Forstner}, {McAuliffe}, {del Rio}, {Berry}, {Aye}, {Stefko}, {de Val-Borro}, {Kulumani}, \& {Murakami}}]{spiceypy}
{Annex}, A., {Pearson}, B., {Seignovert}, B., {et~al.} 2020, \bibinfo{title}{{SpiceyPy: a Pythonic Wrapper for the SPICE Toolkit},} J. Open Source Softw., 5, 2050, \dodoi{10.21105/joss.02050}

\bibitem[{{}{Astropy Collaboration} {et~al.}(2013){Astropy Collaboration}, {Robitaille}, {Tollerud}, {Greenfield}, {Droettboom}, {Bray}, {Aldcroft}, {Davis}, {Ginsburg}, {Price-Whelan}, {Kerzendorf}, {Conley}, {Crighton}, {Barbary}, {Muna}, {Ferguson}, {Grollier}, {Parikh}, {Nair}, {Unther}, {Deil}, {Woillez}, {Conseil}, {Kramer}, {Turner}, {Singer}, {Fox}, {Weaver}, {Zabalza}, {Edwards}, {Azalee Bostroem}, {Burke}, {Casey}, {Crawford}, {Dencheva}, {Ely}, {Jenness}, {Labrie}, {Lim}, {Pierfederici}, {Pontzen}, {Ptak}, {Refsdal}, {Servillat}, \& {Streicher}}]{astropy2013}
{Astropy Collaboration}, {Robitaille}, T.~P., {Tollerud}, E.~J., {et~al.} 2013, \bibinfo{title}{{Astropy: A community Python package for astronomy},} \aap, 558, A33, \dodoi{10.1051/0004-6361/201322068}

\bibitem[{{}{Astropy Collaboration} {et~al.}(2018){Astropy Collaboration}, {Price-Whelan}, {Sip{\H{o}}cz}, {G{\"u}nther}, {Lim}, {Crawford}, {Conseil}, {Shupe}, {Craig}, {Dencheva}, {Ginsburg}, {VanderPlas}, {Bradley}, {P{\'e}rez-Su{\'a}rez}, {de Val-Borro}, {Aldcroft}, {Cruz}, {Robitaille}, {Tollerud}, {Ardelean}, {Babej}, {Bach}, {Bachetti}, {Bakanov}, {Bamford}, {Barentsen}, {Barmby}, {Baumbach}, {Berry}, {Biscani}, {Boquien}, {Bostroem}, {Bouma}, {Brammer}, {Bray}, {Breytenbach}, {Buddelmeijer}, {Burke}, {Calderone}, {Cano Rodr{\'\i}guez}, {Cara}, {Cardoso}, {Cheedella}, {Copin}, {Corrales}, {Crichton}, {D'Avella}, {Deil}, {Depagne}, {Dietrich}, {Donath}, {Droettboom}, {Earl}, {Erben}, {Fabbro}, {Ferreira}, {Finethy}, {Fox}, {Garrison}, {Gibbons}, {Goldstein}, {Gommers}, {Greco}, {Greenfield}, {Groener}, {Grollier}, {Hagen}, {Hirst}, {Homeier}, {Horton}, {Hosseinzadeh}, {Hu}, {Hunkeler}, {Ivezi{\'c}}, {Jain}, {Jenness}, {Kanarek}, {Kendrew}, {Kern}, {Kerzendorf}, {Khvalko}, {King}, {Kirkby}, {Kulkarni},
  {Kumar}, {Lee}, {Lenz}, {Littlefair}, {Ma}, {Macleod}, {Mastropietro}, {McCully}, {Montagnac}, {Morris}, {Mueller}, {Mumford}, {Muna}, {Murphy}, {Nelson}, {Nguyen}, {Ninan}, {N{\"o}the}, {Ogaz}, {Oh}, {Parejko}, {Parley}, {Pascual}, {Patil}, {Patil}, {Plunkett}, {Prochaska}, {Rastogi}, {Reddy Janga}, {Sabater}, {Sakurikar}, {Seifert}, {Sherbert}, {Sherwood-Taylor}, {Shih}, {Sick}, {Silbiger}, {Singanamalla}, {Singer}, {Sladen}, {Sooley}, {Sornarajah}, {Streicher}, {Teuben}, {Thomas}, {Tremblay}, {Turner}, {Terr{\'o}n}, {van Kerkwijk}, {de la Vega}, {Watkins}, {Weaver}, {Whitmore}, {Woillez}, {Zabalza}, \& {Astropy Contributors}}]{astropy2018}
{Astropy Collaboration}, {Price-Whelan}, A.~M., {Sip{\H{o}}cz}, B.~M., {et~al.} 2018, \bibinfo{title}{{The Astropy Project: Building an Open-science Project and Status of the v2.0 Core Package},} \aj, 156, 123, \dodoi{10.3847/1538-3881/aabc4f}

\bibitem[{{}{Astropy Collaboration} {et~al.}(2022){Astropy Collaboration}, {Price-Whelan}, {Lim}, {Earl}, {Starkman}, {Bradley}, {Shupe}, {Patil}, {Corrales}, {Brasseur}, {N{\"o}the}, {Donath}, {Tollerud}, {Morris}, {Ginsburg}, {Vaher}, {Weaver}, {Tocknell}, {Jamieson}, {van Kerkwijk}, {Robitaille}, {Merry}, {Bachetti}, {G{\"u}nther}, {Aldcroft}, {Alvarado-Montes}, {Archibald}, {B{\'o}di}, {Bapat}, {Barentsen}, {Baz{\'a}n}, {Biswas}, {Boquien}, {Burke}, {Cara}, {Cara}, {Conroy}, {Conseil}, {Craig}, {Cross}, {Cruz}, {D'Eugenio}, {Dencheva}, {Devillepoix}, {Dietrich}, {Eigenbrot}, {Erben}, {Ferreira}, {Foreman-Mackey}, {Fox}, {Freij}, {Garg}, {Geda}, {Glattly}, {Gondhalekar}, {Gordon}, {Grant}, {Greenfield}, {Groener}, {Guest}, {Gurovich}, {Handberg}, {Hart}, {Hatfield-Dodds}, {Homeier}, {Hosseinzadeh}, {Jenness}, {Jones}, {Joseph}, {Kalmbach}, {Karamehmetoglu}, {Ka{\l}uszy{\'n}ski}, {Kelley}, {Kern}, {Kerzendorf}, {Koch}, {Kulumani}, {Lee}, {Ly}, {Ma}, {MacBride}, {Maljaars}, {Muna}, {Murphy}, {Norman},
  {O'Steen}, {Oman}, {Pacifici}, {Pascual}, {Pascual-Granado}, {Patil}, {Perren}, {Pickering}, {Rastogi}, {Roulston}, {Ryan}, {Rykoff}, {Sabater}, {Sakurikar}, {Salgado}, {Sanghi}, {Saunders}, {Savchenko}, {Schwardt}, {Seifert-Eckert}, {Shih}, {Jain}, {Shukla}, {Sick}, {Simpson}, {Singanamalla}, {Singer}, {Singhal}, {Sinha}, {Sip{\H{o}}cz}, {Spitler}, {Stansby}, {Streicher}, {{\v{S}}umak}, {Swinbank}, {Taranu}, {Tewary}, {Tremblay}, {de Val-Borro}, {Van Kooten}, {Vasovi{\'c}}, {Verma}, {de Miranda Cardoso}, {Williams}, {Wilson}, {Winkel}, {Wood-Vasey}, {Xue}, {Yoachim}, {Zhang}, {Zonca}, \& {Astropy Project Contributors}}]{astropy2022}
{Astropy Collaboration}, {Price-Whelan}, A.~M., {Lim}, P.~L., {et~al.} 2022, \bibinfo{title}{{The Astropy Project: Sustaining and Growing a Community-oriented Open-source Project and the Latest Major Release (v5.0) of the Core Package},} \apj, 935, 167, \dodoi{10.3847/1538-4357/ac7c74}

\bibitem[{S.~D. Bale {et~al.}(2016)Bale, Goetz, Harvey, Turin, Bonnell, Dudok de Wit, Ergun, MacDowall, Pulupa, Andre, Bolton, Bougeret, Bowen, Burgess, Cattell, Chandran, Chaston, Chen, Choi, Connerney, Cranmer, Diaz-Aguado, Donakowski, Drake, Farrell, Fergeau, Fermin, Fischer, Fox, Glaser, Goldstein, Gordon, Hanson, Harris, Hayes, Hinze, Hollweg, Horbury, Howard, Hoxie, Jannet, Karlsson, Kasper, Kellogg, Kien, Klimchuk, Krasnoselskikh, Krucker, Lynch, Maksimovic, Malaspina, Marker, Martin, Martinez-Oliveros, McCauley, McComas, McDonald, Meyer-Vernet, Moncuquet, Monson, Mozer, Murphy, Odom, Oliverson, Olson, Parker, Pankow, Phan, Quataert, Quinn, Ruplin, Salem, Seitz, Sheppard, Siy, Stevens, Summers, Szabo, Timofeeva, Vaivads, Velli, Yehle, Werthimer, \& Wygant}]{Bale2016}
Bale, S.~D., Goetz, K., Harvey, P.~R., {et~al.} 2016, \bibinfo{title}{The FIELDS Instrument Suite for Solar Probe Plus: Measuring the Coronal Plasma and Magnetic Field, Plasma Waves and Turbulence, and Radio Signatures of Solar Transients,} Space Sci. Rev., 204, 49, \dodoi{10.1007/s11214-016-0244-5}

\bibitem[{E.~G. Blackman {\&} A. Brandenburg(2003)Blackman \& Brandenburg}]{Blackman2003}
Blackman, E.~G., \& Brandenburg, A. 2003, \bibinfo{title}{Doubly Helical Coronal Ejections from Dynamos and Their Role in Sustaining the Solar Cycle,} Astrophys. J., 584, L99, \dodoi{10.1086/368374}

\bibitem[{S. Bourouaine {\&} J.~C. Perez(2018)Bourouaine \& Perez}]{Bourouaine2018}
Bourouaine, S., \& Perez, J.~C. 2018, \bibinfo{title}{On the Limitations of Taylor’s Hypothesis in Parker Solar Probe’s Measurements near the Alfvén Critical Point,} Astrophys. J. Lett., 858, L20, \dodoi{10.3847/2041-8213/aabccf}

\bibitem[{S. Bourouaine {\&} J.~C. Perez(2019)Bourouaine \& Perez}]{Bourouaine2019}
Bourouaine, S., \& Perez, J.~C. 2019, \bibinfo{title}{On the Interpretation of Parker Solar Probe Turbulent Signals,} Astrophys. J. Lett., 879, L16, \dodoi{10.3847/2041-8213/ab288a}

\bibitem[{T.~A. Bowen {et~al.}(2018)Bowen, Badman, Hellinger, \& Bale}]{Bowen2018}
Bowen, T.~A., Badman, S., Hellinger, P., \& Bale, S.~D. 2018, \bibinfo{title}{Density Fluctuations in the Solar Wind Driven by Alfvén Wave Parametric Decay,} Astrophys. J. Lett., 854, L33, \dodoi{10.3847/2041-8213/aaabbe}

\bibitem[{T.~A. Bowen {et~al.}(2020)Bowen, Mallet, Bale, Bonnell, Case, Chandran, Chasapis, Chen, Duan, Dudok~de Wit, Goetz, Halekas, Harvey, Kasper, Korreck, Larson, Livi, MacDowall, Malaspina, McManus, Pulupa, Stevens, \& Whittlesey}]{Bowen2020}
Bowen, T.~A., Mallet, A., Bale, S.~D., {et~al.} 2020, \bibinfo{title}{Constraining Ion-Scale Heating and Spectral Energy Transfer in Observations of Plasma Turbulence,} Phys. Rev. Lett., 125, 025102, \dodoi{10.1103/physrevlett.125.025102}

\bibitem[{A. Brandenburg(2001)Brandenburg}]{Brandenburg2001}
Brandenburg, A. 2001, \bibinfo{title}{The Inverse Cascade and Nonlinear Alpha‐Effect in Simulations of Isotropic Helical Hydromagnetic Turbulence,} Astrophys. J., 550, 824, \dodoi{10.1086/319783}

\bibitem[{A. Brandenburg {et~al.}(2011)Brandenburg, Subramanian, Balogh, \& Goldstein}]{Brandenburg2011}
Brandenburg, A., Subramanian, K., Balogh, A., \& Goldstein, M.~L. 2011, \bibinfo{title}{SCALE DEPENDENCE OF MAGNETIC HELICITY IN THE SOLAR WIND,} Astrophys. J., 734, 9, \dodoi{10.1088/0004-637x/734/1/9}

\bibitem[{M. Brown {et~al.}(1999)Brown, Canfield, Field, Kulsrud, Pevtsov, Rosner, \& Seehaf}]{Brown1999}
Brown, M., Canfield, R., Field, G., {et~al.} 1999, \bibinfo{title}{Magnetic Helicity in Space and Laboratory Plasmas: Editorial Summary,} in Magnetic Helicity in Space and Laboratory Plasmas, ed. M.~R. Brown, R.~C. Canfield, \& A.~A. Pevtsov (Washington DC, US: American Geophysical Union), 301--304

\bibitem[{R. Bruno {\&} V. Carbone(2013)Bruno \& Carbone}]{Bruno2013}
Bruno, R., \& Carbone, V. 2013, \bibinfo{title}{The Solar Wind as a Turbulence Laboratory,} Living Rev. Sol. Phys., 10, 2, \dodoi{10.12942/lrsp-2013-2}

\bibitem[{R. Bruno {\&} M. Dobrowolny(1986)Bruno \& Dobrowolny}]{Bruno1986}
Bruno, R., \& Dobrowolny, M. 1986, \bibinfo{title}{Spectral measurement of magnetic energy and magnetic helicity between 0.29 and 0.97 {AU},} Ann. Geophys. Ser. {A}, 4, 17.
\newblock \url{https://ui.adsabs.harvard.edu/abs/1986AnGeo...4...17B}

\bibitem[{A.~W. Case {et~al.}(2020)Case, Kasper, Stevens, Korreck, Paulson, Daigneau, Caldwell, Freeman, Henry, Klingensmith, Bookbinder, Robinson, Berg, Tiu, Wright, Reinhart, Curtis, Ludlam, Larson, Whittlesey, Livi, Klein, \& Martinović}]{Case2020}
Case, A.~W., Kasper, J.~C., Stevens, M.~L., {et~al.} 2020, \bibinfo{title}{The Solar Probe Cup on the Parker Solar Probe,} Astrophys. J. Suppl. Ser., 246, 43, \dodoi{10.3847/1538-4365/ab5a7b}

\bibitem[{C.~H.~K. Chen {et~al.}(2020)Chen, Bale, Bonnell, Borovikov, Bowen, Burgess, Case, Chandran, de~Wit, Goetz, Harvey, Kasper, Klein, Korreck, Larson, Livi, MacDowall, Malaspina, Mallet, McManus, Moncuquet, Pulupa, Stevens, \& Whittlesey}]{Chen2020}
Chen, C. H.~K., Bale, S.~D., Bonnell, J.~W., {et~al.} 2020, \bibinfo{title}{The Evolution and Role of Solar Wind Turbulence in the Inner Heliosphere,} Astrophys. J. Suppl. Ser., 246, 53, \dodoi{10.3847/1538-4365/ab60a3}

\bibitem[{G. Consolini {et~al.}(2020)Consolini, Alberti, \& Carbone}]{Consolini2020}
Consolini, G., Alberti, T., \& Carbone, V. 2020, \bibinfo{title}{On Yaglom’s Law for the Interplanetary Proton Density and Temperature Fluctuations in Solar Wind Turbulence,} Entropy, 22, 1419, \dodoi{10.3390/e22121419}

\bibitem[{L. Del~Zanna {et~al.}(2001)Del~Zanna, Velli, \& Londrillo}]{DelZanna2001}
Del~Zanna, L., Velli, M., \& Londrillo, P. 2001, \bibinfo{title}{Parametric decay of circularly polarized Alfvén waves: Multidimensional simulations in periodic and open domains,} Astron. Astrophys., 367, 705, \dodoi{10.1051/0004-6361:20000455}

\bibitem[{D. Duan {et~al.}(2021)Duan, He, Bowen, Woodham, Wang, Chen, Mallet, \& Bale}]{Duan2021}
Duan, D., He, J., Bowen, T.~A., {et~al.} 2021, \bibinfo{title}{Anisotropy of Solar Wind Turbulence in the Inner Heliosphere at Kinetic Scales: {PSP} Observations,} Astrophys. J. Lett., 915, L8, \dodoi{10.3847/2041-8213/ac07ac}

\bibitem[{D. {Duan} {et~al.}(2020{\natexlab{a}}){Duan}, {He}, {Wu}, \& {Verscharen}}]{Duan2020_ApJ}
{Duan}, D., {He}, J., {Wu}, H., \& {Verscharen}, D. 2020{\natexlab{a}}, \bibinfo{title}{{Magnetic Energy Transfer and Distribution between Protons and Electrons for Alfv{\'e}nic Waves at Kinetic Scales in Wavenumber Space},} \apj, 896, 47, \dodoi{10.3847/1538-4357/ab8ad2}

\bibitem[{D. {Duan} {et~al.}(2020{\natexlab{b}}){Duan}, {Bowen}, {Chen}, {Mallet}, {He}, {Bale}, {Vech}, {Kasper}, {Pulupa}, {Bonnell}, {Case}, {de Wit}, {Goetz}, {Harvey}, {Korreck}, {Larson}, {Livi}, {MacDowall}, {Malaspina}, {Stevens}, \& {Whittlesey}}]{Duan2020_ApJS}
{Duan}, D., {Bowen}, T.~A., {Chen}, C. H.~K., {et~al.} 2020{\natexlab{b}}, \bibinfo{title}{{The Radial Dependence of Proton-scale Magnetic Spectral Break in Slow Solar Wind during PSP Encounter 2},} \apjs, 246, 55, \dodoi{10.3847/1538-4365/ab672d}

\bibitem[{N.~J. Fox {et~al.}(2016)Fox, Velli, Bale, Decker, Driesman, Howard, Kasper, Kinnison, Kusterer, Lario, Lockwood, McComas, Raouafi, \& Szabo}]{Fox2016}
Fox, N.~J., Velli, M.~C., Bale, S.~D., {et~al.} 2016, \bibinfo{title}{The Solar Probe Plus Mission: Humanity’s First Visit to Our Star,} Space Sci. Rev., 204, 7, \dodoi{10.1007/s11214-015-0211-6}

\bibitem[{C. Froment {et~al.}(2023)Froment, Agapitov, Krasnoselskikh, Karbashewski, Dudok~de Wit, Larosa, Colomban, Malaspina, Kretzschmar, Jagarlamudi, Bale, Bonnell, Mozer, \& Pulupa}]{Froment2023}
Froment, C., Agapitov, O.~V., Krasnoselskikh, V., {et~al.} 2023, \bibinfo{title}{Whistler waves generated inside magnetic dips in the young solar wind: Observations of the search-coil magnetometer on board Parker Solar Probe,} Astron. Astrophys., 672, A135, \dodoi{10.1051/0004-6361/202245140}

\bibitem[{K.-H. Glassmeier {et~al.}(2001)Glassmeier, Motschmann, Dunlop, Balogh, Acuña, Carr, Musmann, Forna\c{c}on, Schweda, Vogt, Georgescu, \& Buchert}]{Glassmeier2001}
Glassmeier, K.-H., Motschmann, U., Dunlop, M., {et~al.} 2001, \bibinfo{title}{Cluster as a wave telescope --- first results from the fluxgate magnetometer,} Ann. Geophys., 19, 1439, \dodoi{10.5194/angeo-19-1439-2001}

\bibitem[{M.~L. Goldstein {\&} D.~A. Roberts(1999)Goldstein \& Roberts}]{Goldstein1999}
Goldstein, M.~L., \& Roberts, D.~A. 1999, \bibinfo{title}{Magnetohydrodynamic turbulence in the solar wind,} Phys. Plasmas, 6, 4154, \dodoi{10.1063/1.873680}

\bibitem[{R. Grappin {\&} M. Velli(1996)Grappin \& Velli}]{Grappin1996}
Grappin, R., \& Velli, M. 1996, \bibinfo{title}{Waves and streams in the expanding solar wind,} J. Geophys. Res. Space Phys., 101, 425, \dodoi{10.1029/95ja02147}

\bibitem[{C.~R. {Harris} {et~al.}(2020){Harris}, {Millman}, {van der Walt}, {Gommers}, {Virtanen}, {Cournapeau}, {Wieser}, {Taylor}, {Berg}, {Smith}, {Kern}, {Picus}, {Hoyer}, {van Kerkwijk}, {Brett}, {Haldane}, {del R{\'\i}o}, {Wiebe}, {Peterson}, {G{\'e}rard-Marchant}, {Sheppard}, {Reddy}, {Weckesser}, {Abbasi}, {Gohlke}, \& {Oliphant}}]{numpy}
{Harris}, C.~R., {Millman}, K.~J., {van der Walt}, S.~J., {et~al.} 2020, \bibinfo{title}{{Array programming with NumPy},} \nat, 585, 357, \dodoi{10.1038/s41586-020-2649-2}

\bibitem[{D.~P. Hartley {et~al.}(2017)Hartley, Kletzing, Kurth, Hospodarsky, Bounds, Averkamp, Bonnell, Santolík, \& Wygant}]{Hartley2017}
Hartley, D.~P., Kletzing, C.~A., Kurth, W.~S., {et~al.} 2017, \bibinfo{title}{An improved sheath impedance model for the Van Allen Probes EFW instrument: Effects of the spin axis antenna,} J. Geophys. Res. Space Phys., 122, 4420, \dodoi{10.1002/2016ja023597}

\bibitem[{J. He {et~al.}(2011)He, Marsch, Tu, Yao, \& Tian}]{He2011}
He, J., Marsch, E., Tu, C., Yao, S., \& Tian, H. 2011, \bibinfo{title}{POSSIBLE EVIDENCE OF ALFVÉN-CYCLOTRON WAVES IN THE ANGLE DISTRIBUTION OF MAGNETIC HELICITY OF SOLAR WIND TURBULENCE,} Astrophys. J., 731, 85, \dodoi{10.1088/0004-637x/731/2/85}

\bibitem[{J. {He} {et~al.}(2020){He}, {Zhu}, {Verscharen}, {Duan}, {Zhao}, \& {Wang}}]{He2020}
{He}, J., {Zhu}, X., {Verscharen}, D., {et~al.} 2020, \bibinfo{title}{{Spectra of Diffusion, Dispersion, and Dissipation for Kinetic Alfv{\'e}nic and Compressive Turbulence: Comparison between Kinetic Theory and Measurements from MMS},} \apj, 898, 43, \dodoi{10.3847/1538-4357/ab9174}

\bibitem[{S.~Y. Huang {et~al.}(2020)Huang, Zhang, Sahraoui, He, Yuan, Andrés, Hadid, Deng, Jiang, Yu, Xiong, Wei, Xu, Bale, \& Kasper}]{Huang2020}
Huang, S.~Y., Zhang, J., Sahraoui, F., {et~al.} 2020, \bibinfo{title}{Kinetic Scale Slow Solar Wind Turbulence in the Inner Heliosphere: Coexistence of Kinetic Alfvén Waves and Alfvén Ion Cyclotron Waves,} Astrophys. J. Lett., 897, L3, \dodoi{10.3847/2041-8213/ab9abb}

\bibitem[{J.~D. {Hunter}(2007){Hunter}}]{matplotlib}
{Hunter}, J.~D. 2007, \bibinfo{title}{{Matplotlib: A 2D Graphics Environment},} Comput. Sci. Eng., 9, 90, \dodoi{10.1109/MCSE.2007.55}

\bibitem[{T. Imachi {et~al.}(2007)Imachi, Yagitani, Higashi, \& Nagano}]{Imachi2007}
Imachi, T., Yagitani, S., Higashi, R., \& Nagano, I. 2007, \bibinfo{title}{Rheometry experiment for a wire antenna aboard spacecraft at low frequencies,} Electron. Commun. Jpn. Pt. I Commun., 90, 45, \dodoi{10.1002/ecja.20353}

\bibitem[{V.~K. Jagarlamudi {et~al.}(2021)Jagarlamudi, Dudok~de Wit, Froment, Krasnoselskikh, Larosa, Bercic, Agapitov, Halekas, Kretzschmar, Malaspina, Moncuquet, Bale, Case, Kasper, Korreck, Larson, Pulupa, Stevens, \& Whittlesey}]{Jagarlamudi2021}
Jagarlamudi, V.~K., Dudok~de Wit, T., Froment, C., {et~al.} 2021, \bibinfo{title}{Whistler wave occurrence and the interaction with strahl electrons during the first encounter of Parker Solar Probe,} Astron. Astrophys., 650, A9, \dodoi{10.1051/0004-6361/202039808}

\bibitem[{S. Karbashewski {et~al.}(2023)Karbashewski, Agapitov, Kim, Mozer, Bonnell, Froment, de~Wit, Bale, Malaspina, \& Raouafi}]{Karbashewski2023}
Karbashewski, S., Agapitov, O.~V., Kim, H.~Y., {et~al.} 2023, \bibinfo{title}{Whistler Wave Observations by Parker Solar Probe During Encounter 1: Counter-propagating Whistlers Collocated with Magnetic Field Inhomogeneities and their Application to Electric Field Measurement Calibration,} Astrophys. J., 947, 73, \dodoi{10.3847/1538-4357/acc527}

\bibitem[{H. Kim {\&} J. Cho(2015)Kim \& Cho}]{Kim2015}
Kim, H., \& Cho, J. 2015, \bibinfo{title}{INVERSE CASCADE IN IMBALANCED ELECTRON MAGNETOHYDRODYNAMIC TURBULENCE,} Astrophys. J., 801, 75, \dodoi{10.1088/0004-637x/801/2/75}

\bibitem[{K.~G. Klein {et~al.}(2014)Klein, Howes, \& TenBarge}]{Klein2014}
Klein, K.~G., Howes, G.~G., \& TenBarge, J.~M. 2014, \bibinfo{title}{THE VIOLATION OF THE TAYLOR HYPOTHESIS IN MEASUREMENTS OF SOLAR WIND TURBULENCE,} Astrophys. J. Lett., 790, L20, \dodoi{10.1088/2041-8205/790/2/l20}

\bibitem[{R.~J. Leamon {et~al.}(1998)Leamon, Smith, Ness, Matthaeus, \& Wong}]{Leamon1998}
Leamon, R.~J., Smith, C.~W., Ness, N.~F., Matthaeus, W.~H., \& Wong, H.~K. 1998, \bibinfo{title}{Observational constraints on the dynamics of the interplanetary magnetic field dissipation range,} J. Geophys. Res. Space Phys., 103, 4775, \dodoi{10.1029/97ja03394}

\bibitem[{R. Livi {et~al.}(2022)Livi, Larson, Kasper, Abiad, Case, Klein, Curtis, Dalton, Stevens, Korreck, Ho, Robinson, Tiu, Whittlesey, Verniero, Halekas, McFadden, Marckwordt, Slagle, Abatcha, Rahmati, \& McManus}]{Livi2022}
Livi, R., Larson, D.~E., Kasper, J.~C., {et~al.} 2022, \bibinfo{title}{The {Solar} {Probe} {ANalyzer}---{Ions} on the Parker Solar Probe,} Astrophys. J., 938, 138, \dodoi{10.3847/1538-4357/ac93f5}

\bibitem[{E. Marsch {\&} C.~Y. Tu(1993)Marsch \& Tu}]{Marsch1993}
Marsch, E., \& Tu, C.~Y. 1993, \bibinfo{title}{Modeling results on spatial transport and spectral transfer of solar wind Alfvénic turbulence,} J. Geophys. Res. Space Phys., 98, 21045, \dodoi{10.1029/93ja02365}

\bibitem[{W.~H. Matthaeus {\&} M.~L. Goldstein(1982)Matthaeus \& Goldstein}]{Matthaeus1982_jgr}
Matthaeus, W.~H., \& Goldstein, M.~L. 1982, \bibinfo{title}{Measurement of the rugged invariants of magnetohydrodynamic turbulence in the solar wind,} J. Geophys. Res. Space Phys., 87, 6011, \dodoi{10.1029/ja087ia08p06011}

\bibitem[{W.~H. Matthaeus {et~al.}(1982)Matthaeus, Goldstein, \& Smith}]{Matthaeus1982_prl}
Matthaeus, W.~H., Goldstein, M.~L., \& Smith, C. 1982, \bibinfo{title}{Evaluation of Magnetic Helicity in Homogeneous Turbulence,} Phys. Rev. Lett., 48, 1256, \dodoi{10.1103/physrevlett.48.1256}

\bibitem[{R. Meyrand {et~al.}(2021)Meyrand, Squire, Schekochihin, \& Dorland}]{Meyrand2021}
Meyrand, R., Squire, J., Schekochihin, A., \& Dorland, W. 2021, \bibinfo{title}{On the violation of the zeroth law of turbulence in space plasmas,} J. Plasma Phys., 87, 535870301, \dodoi{10.1017/s0022377821000489}

\bibitem[{W.-C. M\"{u}ller {\&} S.~K. Malapaka(2012)M\"{u}ller \& Malapaka}]{Mueller2012}
M\"{u}ller, W.-C., \& Malapaka, S.~K. 2012, \bibinfo{title}{Role of helicities for the dynamics of turbulent magnetic fields,} Geophys. Astrophys. Fluid Dyn., 107, 93, \dodoi{10.1080/03091929.2012.688292}

\bibitem[{S.~J. {Mumford} {et~al.}(2025){Mumford}, {Freij}, {Stansby}, {Christe}, {Shih}, {Ireland}, {Mayer}, {Hughitt}, {Ryan}, {Liedtke}, {Barnes}, {Hayes}, {P{\'e}rez-Su{\'a}rez}, {I.}, {Chakraborty}, {Inglis}, {Pattnaik}, {Sip{\H{o}}cz}, {Hossam}, {MacBride}, {Sharma}, {Leonard}, {Hewett}, {Hamilton}, {Manhas}, {Panda}, {Earnshaw}, {Choudhary}, {Kumar}, {Singh}, {Chanda}, {Maloney}, {Akramul Haque}, {Hole}, {Wilson}, {Gilly}, {Kirk}, {Mueller}, {Konge}, {Wentzel-Long}, {Srivastava}, {Bennett}, {Jain}, {Zivadinovic}, {Baruah}, {Arbolante}, {Simon}, {Charlton}, {Mishra}, {Paul}, {Verma}, {Chorley}, {Chouhan}, {Himanshu}, {Mason}, {Modi}, {Sharma}, {Tyagi}, {Sinha}, {Naman9639}, {Bobra}, {Campos Rozo}, {Manley}, {Ivashkiv}, {Laitinen}, {Chatterjee}, {Dixit}, {Gieseler}, {Dulange}, {von Forstner}, {Baz{\'a}n}, {Akira Stern}, {Shukla}, {Evans}, {Jain}, {Malocha}, {Ghosh}, {Airmansmith97}, {Khushwaha}, {Sta{\'n}czak}, {Singh}, {Ranjan Singh}, {De Visscher}, {Verma}, {Lemos}, {Agrawal}, {Alam}, {Graham},
  {Buddhika}, {Collier}, {Pathak}, {Rideout}, {Roy}, {Sharma}, {Mo}, {Garcia Briseno}, {Shah}, {Park}, {Bates}, {Shukla}, {Giger}, {Mishra}, {Yelgoe}, {Sharma}, {Goel}, {Taylor}, {Cetusic}, {Reiter}, {Jacob}, {Inchaurrandieta}, {Sharma}, {Dacie}, {Dubey}, {Eigenbrot}, {Mampaey}, {Bray}, {Murphy}, {Surve}, {Zahniy}, {Sidhu}, {Meszaros}, {Parkhi}, {Russell}, {Bose}, {Pandey}, {Mahapatra}, {Price-Whelan}, {Jahagirdar}, {Chicrala}, {Mishra}, {Ankit}, {Guennou}, {D'Avella}, {Williams}, {Verma}, {Ballew}, {Agrawal}, {Manasia}, {Kulkarni}, {Singh}, {Streicher}, {Lodha}, {Van Kooten}, {Mishra}, {Robitaille}, {Augspurger}, {Krishan}, {Jeon}, {Bahuleyan}, {Pimparkar}, {Bhope}, {Saeed}, {Singh Gaba}, {Hill}, {Wiedemann}, {Molina}, {Khetarpal}, {Ke{\c{s}}kek}, {Habib}, {Letts}, {Singaravelan}, {Ranjan}, {Pandey}, {Altunian}, {Gomillion}, {Agarwal}, {Kothari}, {Malik}, {Nomiya}, {Burnett}, {Stevens}, {Shauryam}, {Kaszynski}, {Wang}, {Mehrotra}, {Tang}, {Sinha}, {Smith}, {Kustov}, {Bastola}, {Stone}, {Bard}, {Lowder},
  {Robert}, {Behn}, {Mansky}, {Arias}, {Paganin}, {Tollerud}, {Mackenzie Dover}, {Verstringe}, {FreyaJain}, {Kdimati}, {Kumar}, {Hardik}, {Mathur}, {Babuschkin}, {Calixto}, {Wimbish}, \& {Qing}}]{sunpy_zenodo}
{Mumford}, S.~J., {Freij}, N., {Stansby}, D., {et~al.} 2025, {sunpy: A Core Package for Solar Physics}, v7.0.1 Zenodo, \dodoi{10.5281/zenodo.16638197}

\bibitem[{Y. Narita(2023)Narita}]{Narita2023}
Narita, Y. 2023, \bibinfo{title}{Helicities in Geophysics, Astrophysics and Beyond,} in Geophysical Monograph Series, Vol. 283, Helicities in Geophysics, Astrophysics and Beyond, ed. K.~Kuzanyan, N.~Yokoi, M.~K. Georgoulis, \& R.~Stepanov (New Jersey, USA: John Wiley \& Sons, Inc.), 105--116, \dodoi{10.1002/9781119841715.ch7}

\bibitem[{Y. Narita {et~al.}(2009)Narita, Kleindienst, \& Glassmeier}]{Narita2009}
Narita, Y., Kleindienst, G., \& Glassmeier, K.-H. 2009, \bibinfo{title}{Evaluation of magnetic helicity density in the wave number domain using multi-point measurements in space,} Ann. Geophys., 27, 3967, \dodoi{10.5194/angeo-27-3967-2009}

\bibitem[{J.~C. Perez {et~al.}(2021)Perez, Bourouaine, Chen, \& Raouafi}]{Perez2021}
Perez, J.~C., Bourouaine, S., Chen, C. H.~K., \& Raouafi, N.~E. 2021, \bibinfo{title}{Applicability of Taylor’s hypothesis during Parker Solar Probe perihelia,} Astron. Astrophys., 650, A22, \dodoi{10.1051/0004-6361/202039879}

\bibitem[{D. Perrone {et~al.}(2022)Perrone, Perri, Bruno, Stansby, D’Amicis, Jagarlamudi, Laker, Toledo-Redondo, Stawarz, Telloni, De~Marco, Owen, Raines, Settino, Lavraud, Maksimovic, Vaivads, Phan, Fargette, Louarn, \& Zouganelis}]{Perrone2022}
Perrone, D., Perri, S., Bruno, R., {et~al.} 2022, \bibinfo{title}{Evolution of coronal hole solar wind in the inner heliosphere: Combined observations by Solar Orbiter and Parker Solar Probe,} Astron. Astrophys., 668, A189, \dodoi{10.1051/0004-6361/202243989}

\bibitem[{J.~J. Podesta(2013)Podesta}]{Podesta2013}
Podesta, J.~J. 2013, \bibinfo{title}{Evidence of Kinetic Alfv\'{e}n Waves in the Solar Wind at 1 {AU},} Sol. Phys., 286, 529, \dodoi{10.1007/s11207-013-0258-z}

\bibitem[{J.~J. Podesta {\&} S.~P. Gary(2011)Podesta \& Gary}]{Podesta2011}
Podesta, J.~J., \& Gary, S.~P. 2011, \bibinfo{title}{MAGNETIC HELICITY SPECTRUM OF SOLAR WIND FLUCTUATIONS AS A FUNCTION OF THE ANGLE WITH RESPECT TO THE LOCAL MEAN MAGNETIC FIELD,} Astrophys. J., 734, 15, \dodoi{10.1088/0004-637x/734/1/15}

\bibitem[{D. Ram\'{i}rez {et~al.}(2022)Ram\'{i}rez, Santamar\'{i}a, \& Scharf}]{Ramrez2022}
Ram\'{i}rez, D., Santamar\'{i}a, I., \& Scharf, L. 2022, Coherence: In Signal Processing and Machine Learning (Springer International Publishing), \dodoi{10.1007/978-3-031-13331-2}

\bibitem[{C.~W. Smith(2003)Smith}]{Smith2003}
Smith, C.~W. 2003, \bibinfo{title}{Magnetic helicity in the solar wind,} Adv. Space Res., 32, 1971, \dodoi{10.1016/s0273-1177(03)90635-1}

\bibitem[{L. Sorriso-Valvo {et~al.}(2023)Sorriso-Valvo, Marino, Foldes, Lév\^eque, D’Amicis, Bruno, Telloni, \& Yordanova}]{SorrisoValvo2023}
Sorriso-Valvo, L., Marino, R., Foldes, R., {et~al.} 2023, \bibinfo{title}{Helios 2 observations of solar wind turbulence decay in the inner heliosphere,} Astron. Astrophys., 672, A13, \dodoi{10.1051/0004-6361/202244889}

\bibitem[{L. Sorriso-Valvo {et~al.}(2007)Sorriso-Valvo, Marino, Carbone, Noullez, Lepreti, Veltri, Bruno, Bavassano, \& Pietropaolo}]{SorrisoValvo2007}
Sorriso-Valvo, L., Marino, R., Carbone, V., {et~al.} 2007, \bibinfo{title}{Observation of Inertial Energy Cascade in Interplanetary Space Plasma,} Phys. Rev. Lett., 99, 115001, \dodoi{10.1103/physrevlett.99.115001}

\bibitem[{J. Squire {et~al.}(2022)Squire, Meyrand, Kunz, Arzamasskiy, Schekochihin, \& Quataert}]{Squire2022}
Squire, J., Meyrand, R., Kunz, M.~W., {et~al.} 2022, \bibinfo{title}{High-frequency heating of the solar wind triggered by low-frequency turbulence,} Nat. Astron., 6, 715, \dodoi{10.1038/s41550-022-01624-z}

\bibitem[{{}{SunPy Community} {et~al.}(2020){SunPy Community}, {Barnes}, {Bobra}, {Christe}, {Freij}, {Hayes}, {Ireland}, {Mumford}, {Perez-Suarez}, {Ryan}, {Shih}, {Chanda}, {Glogowski}, {Hewett}, {Hughitt}, {Hill}, {Hiware}, {Inglis}, {Kirk}, {Konge}, {Mason}, {Maloney}, {Murray}, {Panda}, {Park}, {Pereira}, {Reardon}, {Savage}, {Sip{\H{o}}cz}, {Stansby}, {Jain}, {Taylor}, {Yadav}, {Rajul}, \& {Dang}}]{sunpy2020}
{SunPy Community}, {Barnes}, W.~T., {Bobra}, M.~G., {et~al.} 2020, \bibinfo{title}{{The SunPy Project: Open Source Development and Status of the Version 1.0 Core Package},} \apj, 890, 68, \dodoi{10.3847/1538-4357/ab4f7a}

\bibitem[{G.~I. Taylor(1938)Taylor}]{Taylor1938}
Taylor, G.~I. 1938, \bibinfo{title}{The Spectrum of Turbulence,} Proc. R. Soc. A, 164, 476, \dodoi{10.1098/rspa.1938.0032}

\bibitem[{D. Telloni {et~al.}(2015)Telloni, Bruno, \& Trenchi}]{Telloni2015}
Telloni, D., Bruno, R., \& Trenchi, L. 2015, \bibinfo{title}{RADIAL EVOLUTION OF SPECTRAL CHARACTERISTICS OF MAGNETIC FIELD FLUCTUATIONS AT PROTON SCALES,} Astrophys. J., 805, 46, \dodoi{10.1088/0004-637x/805/1/46}

\bibitem[{A. Tenerani {\&} M. Velli(2013)Tenerani \& Velli}]{Tenerani2013}
Tenerani, A., \& Velli, M. 2013, \bibinfo{title}{Parametric decay of radial Alfvén waves in the expanding accelerating solar wind,} J. Geophys. Res. Space Phys., 118, 7507, \dodoi{10.1002/2013ja019293}

\bibitem[{C.~Y. Tu {\&} E. Marsch(1993)Tu \& Marsch}]{Tu1993}
Tu, C.~Y., \& Marsch, E. 1993, \bibinfo{title}{A model of solar wind fluctuations with two components: Alfvén waves and convective structures,} J. Geophys. Res. Space Phys., 98, 1257, \dodoi{10.1029/92ja01947}

\bibitem[{R. Vink {et~al.}(2025)Vink, de~Gooijer, Beedie, Gorelli, {nameexhaustion}, Peters, Burghoorn, Guo, van Zundert, {Marshall}, Hulselmans, Grinstead, Manley, {chielP}, Turner-Trauring, Mitchell, Santamaria, Harbeck, Heres, {eitsupi}, Magarick, {deanm}, Genockey, Koutsouris, {Robin}, Wilksch, Brannigan, Leitao, \& van Gelderen}]{polars}
Vink, R., de~Gooijer, S., Beedie, A., {et~al.} 2025, pola-rs/polars: Python Polars 1.27.1, Zenodo, \dodoi{10.5281/ZENODO.15194971}

\bibitem[{P. {Virtanen} {et~al.}(2020){Virtanen}, {Gommers}, {Oliphant}, {Haberland}, {Reddy}, {Cournapeau}, {Burovski}, {Peterson}, {Weckesser}, {Bright}, {van der Walt}, {Brett}, {Wilson}, {Millman}, {Mayorov}, {Nelson}, {Jones}, {Kern}, {Larson}, {Carey}, {Polat}, {Feng}, {Moore}, {VanderPlas}, {Laxalde}, {Perktold}, {Cimrman}, {Henriksen}, {Quintero}, {Harris}, {Archibald}, {Ribeiro}, {Pedregosa}, {van Mulbregt}, \& {SciPy 1. 0 Contributors}}]{scipy}
{Virtanen}, P., {Gommers}, R., {Oliphant}, T.~E., {et~al.} 2020, \bibinfo{title}{{SciPy 1.0: fundamental algorithms for scientific computing in Python},} Nat. Med., 17, 261, \dodoi{10.1038/s41592-019-0686-2}

\bibitem[{Y. Wang {\&} O. Chkhetiani(2023{\natexlab{a}})Wang \& Chkhetiani}]{Wang2023}
Wang, Y., \& Chkhetiani, O. 2023{\natexlab{a}}, \bibinfo{title}{Four-thirds law of energy and magnetic helicity in electron and Hall magnetohydrodynamic fluids,} Phys. D, 454, 133835, \dodoi{10.1016/j.physd.2023.133835}

\bibitem[{P. Welch(1967)Welch}]{Welch1967}
Welch, P. 1967, \bibinfo{title}{The use of fast Fourier transform for the estimation of power spectra: A method based on time averaging over short, modified periodograms,} IEEE Trans. Audio Electroacoustics, 15, 70, \dodoi{10.1109/tau.1967.1161901}

\bibitem[{J.~M. Wilcox {\&} N.~F. Ness(1965)Wilcox \& Ness}]{Wilcox1965}
Wilcox, J.~M., \& Ness, N.~F. 1965, \bibinfo{title}{Quasi-stationary corotating structure in the interplanetary medium,} J. Geophys. Res., 70, 5793, \dodoi{10.1029/jz070i023p05793}

\bibitem[{M. Wilczek {\&} Y. Narita(2012)Wilczek \& Narita}]{Wilczek2012}
Wilczek, M., \& Narita, Y. 2012, \bibinfo{title}{Wave-number–frequency spectrum for turbulence from a random sweeping hypothesis with mean flow,} Phys. Rev. E, 86, 066308, \dodoi{10.1103/physreve.86.066308}

\bibitem[{L.~B. Wilson {et~al.}(2022)Wilson, Goodrich, Turner, Cohen, Whittlesey, \& Schwartz}]{Wilson2022}
Wilson, L.~B., Goodrich, K.~A., Turner, D.~L., {et~al.} 2022, \bibinfo{title}{The need for accurate measurements of thermal velocity distribution functions in the solar wind,} Front. Astron. Space Sci., 9, 1063841, \dodoi{10.3389/fspas.2022.1063841}

\bibitem[{H. Wu {et~al.}(2022)Wu, Tu, He, Wang, \& Yang}]{Wu2022}
Wu, H., Tu, C., He, J., Wang, X., \& Yang, L. 2022, \bibinfo{title}{The Yaglom Scaling of the Third-order Structure Functions in the Inner Heliosphere Observed by Helios 1 and 2,} Astrophys. J., 927, 113, \dodoi{10.3847/1538-4357/ac4fcc}

\bibitem[{H. Wu {et~al.}(2023)Wu, Tu, He, Wang, \& Yang}]{Wu2023}
Wu, H., Tu, C., He, J., Wang, X., \& Yang, L. 2023, \bibinfo{title}{Energy transfer of the solar wind turbulence based on Parker solar probe and other spacecraft observations,} Phys. Plasmas, 30, 020501, \dodoi{10.1063/5.0121140}

\bibitem[{H. {Xie} {et~al.}(2026){Xie}, {He}, {Peng}, \& {Wu}}]{Xie2026}
{Xie}, H., {He}, J., {Peng}, J., \& {Wu}, Z. 2026, \bibinfo{title}{{Energy Redistribution of Ion-scale Turbulence from Quasi-perpendicular to Quasi-parallel Wavevectors as Induced by Field-particle Interactions},} \apjl, 1006, L53, \dodoi{10.3847/2041-8213/ae84c8}

\bibitem[{N. Yokoi {\&} F. Hamba(2007)Yokoi \& Hamba}]{Yokoi2007}
Yokoi, N., \& Hamba, F. 2007, \bibinfo{title}{An application of the turbulent magnetohydrodynamic residual-energy equation model to the solar wind,} Phys. Plasmas, 14, 112904, \dodoi{10.1063/1.2792337}

\end{thebibliography}
\end{document}